# Magnetoresistance effect based on spin-selective transport in nanodevices using chiral molecules

Mizuki Matsuzaka,[a] Kotaro Kashima,[a] Koki Terai,[a] Takumi Ueda,[a] Ryunosuke Miyamoto,[a] Takashi Yamamoto,[a] Kohei Sambe,[b] Tomoyuki Akutagawa[b] and Hideo Kaiju[*ac]

[a]*Faculty of Science and Technology, Keio University, Yokohama, Kanagawa 223-8522, Japan*
[b]*Institute of Multidisciplinary Research for Advanced Materials, Tohoku University, Sendai, Miyagi 980-8577, Japan*
[c]*Center for Spintronics Research Network, Keio University, Yokohama, Kanagawa 223-8522, Japan*
[*]*E-mail: kaiju@appi.keio.ac.jp*

**Abstract**

Recently, chirality-induced spin selectivity (CISS) has been observed in chiral molecules and is attractive for application in magnetoresistance (MR) devices. In this study, we fabricate CISS-based nanodevices consisting of chiral molecules sandwiched between $Ni_{78}Fe_{22}$ and Au electrodes. Prior to device fabrication, we have synthesized the chiral molecule *N*-(3*S*)-3,7-dimethyloctyl[1]benzothieno[3,2-*b*]benzothiophene-2-carboxyamide (*S*-BTBT-CONHR) and established a method for fabricating nanodevice electrodes. We have successfully observed a high degree of spin selectivity in *S*-BTBT-CONHR thin films using magnetic conductive atomic force microscopy (mc-AFM). By combining chiral molecules with our advanced nanofabrication technique, we have successfully fabricated Au/*S*-BTBT-CONHR/$Ni_{78}Fe_{22}$ nanodevices and observed the MR effect in the fabricated devices under a low magnetic field at room temperature. These MR curves correspond to the magnetization states of the $Ni_{78}Fe_{22}$ electrode, indicating that the CISS-based MR effect is successfully observed in the nanodevices under a low magnetic field. This study can lead to the development of CISS-based MR devices under low magnetic fields and provide new insights into the CISS effect mechanism on devices.

## Introduction

Molecular spintronics is an emerging research field that combines spintronics and molecular electronics. Organic molecules are primarily composed of light elements that contribute to weak spin–orbit coupling. Therefore, molecular spintronic devices can exhibit longer spin diffusion lengths than inorganic spintronic devices such as magnetic tunnel junctions (MTJs),[1–6] which consist of an inorganic insulator sandwiched between two magnetic layers. Here, organic molecules are useful as spacers between the two magnetic layers instead of as inorganic insulators in MTJs. Devices consisting of a molecular layer sandwiched between two magnetic layers are called organic spin valves (OSVs), which exhibit the magnetoresistance (MR) effect.[7–14]

Previous studies have demonstrated that nanoscale OSVs can provide better performance than micro- or milli-scale OSVs. For example, an MR ratio of 40% was observed at 11 K in Co/tris-(8-hydroxy-quinoline)aluminum ($Alq_3$)/$La_{0.67}Sr_{0.33}MnO_3$ (LSMO) milli-scale devices,[7] whereas a high MR ratio of 300% was attained in Co/$Alq_3$/LSMO nanojunctions at 2 K because of effective orbital hybridization at the metal–molecule interface.[8] In a theoretical study, a Ni/single 1,4-3-phenyl-dithiolate (1,4-tricene-dithiolate)/Ni junction exhibited an MR of 600%.[15] However, most of the studies on nanoscale devices have been limited to low-temperature experiments.[8,10] In our previous study focusing on the high-mobility molecule 2,7-dioctyl[1]benzothieno[3,2-b][1]benzothiophene (C8-BTBT),[16] we have successfully fabricated $Ni_{78}Fe_{22}$/C8-BTBT/$Ni_{78}Fe_{22}$ nanojunctions by a new method utilizing $Ni_{78}Fe_{22}$ thin-film edges.[9] In these devices, the MR effect has been clearly observed in the BTBT-based molecular nanojunctions at room temperature. These results indicate that our method can be applied to the fabrication of nanojunctions consisting of various types of molecules to investigate their spin transport properties in nanoscale systems.

Recently, chirality-induced spin selectivity (CISS) has been observed in chiral molecules.[17] Chiral molecules exhibit high spin selectivity at room temperature, as reported by magnetic conductive atomic force microscopy (mc-AFM).[18–30] In these studies, a magnetic cantilever was magnetized along the up or down direction and placed in contact with the surface of chiral molecular thin films to measure the current–voltage (*I*–*V*) curves at different positions. The degree of spin selectivity is usually expressed as spin polarization, SP = ($I_{down}$ − $I_{up}$)/($I_{down}$ + $I_{up}$), where $I_{up(down)}$ is the electrical current measured with a cantilever magnetized along the up (down) direction in mc-AFM setup. As the recent study pointed out,[31] it should be noted that SP is different from the well-known spin polarization $P$, which is defined as $P = (D_+ − D_−)/(D_+ + D_−)$, where $D_{+(−)}$ is the density of states of majority (minority) spins at the Fermi energy in magnetic layers and often used in spintronic devices such as MTJs. Therefore, we herein define spin selectivity in CISS, $SS_{CISS} \equiv (I_{down} − I_{up})/(I_{down} + I_{up})$ instead of SP to avoid misunderstanding about spin polarization. In mc-AFM studies, chiral molecules have revealed high $SS_{CISS}$ values, such as approximately 0.3 in enantiopure thiophene derivative 3,3'-bibenzothiophene core functionalized with 2,2'-bithiophene wings,[21] and approximately 0.7 in double-stranded DNA oligomers[19] as well as $C_2$-chiral bifacial indacenodithiophene.[24] As the recent studies claimed, while the mechanism of high $SS_{CISS}$ is under discussion, such as electrostatic barrier modulation,[32,33] $SS_{CISS}$ far exceeds the spin polarization $P$ in the ferromagnetic materials. Because chiral molecules exhibit a high degree of spin selectivity, they are expected to be applied in MR devices. Previous studies have demonstrated MR devices using chiral molecules, which are referred to as CISS-based OSVs. They consist of a chiral molecular layer sandwiched between magnetic and non-magnetic electrodes.[20,22,23,25–29,34–36] Because the chiral molecular layer can act as a spin filter, CISS-based OSVs have only one magnetic layer compared to conventional OSVs, which consist of two magnetic layers. Although the CISS-based OSVs exhibited the MR effect at room temperature, their MR ratios were lower than those estimated using mc-AFM.[20,22,23,25–29] For example, MR ratios were approximately 1%, 2%, and 0.05% in the devices, whereas $SS_{CISS}$ values were approximately 0.5, 0.4, and 0.9 by mc-AFM obtained in helicenes, overcrowded alkenes, and helical π-conjugated materials based on supramolecular nanofibers, respectively.[20,23,26] These studies suggest that the low MR ratio was caused by leakage current through pinholes in the chiral molecules. The contact area between the magnetic cantilever and the chiral molecules was nanoscale in the mc-AFM setup, whereas the junction area of the MR devices was micro- or milli-scale. When electrons pass through pinholes in the chiral molecular layer, they are not spin-polarized, resulting in a reduction in the MR ratio. In this case, because the nanosized contact area could contain no pinholes or fewer pinholes, the degree of spin selectivity obtained by mc-AFM could be higher than that in chiral molecule-based MR devices. However, nanoscale MR devices based on chiral molecules have not yet been fabricated or investigated. Therefore, the mechanism underlying the low MR ratio in chiral molecule-based MR devices has not yet been clarified.

In this study, we focused on fabricating nanoscale MR devices based on chiral molecules. In a previous study, we established a technique for fabricating $Ni_{78}Fe_{22}$ electrodes for molecular nanojunctions.[9] In addition to the $Ni_{78}Fe_{22}$ electrode as a magnetic electrode, a nonmagnetic electrode and a chiral molecule are necessary for the fabrication of chiral molecule-based MR devices. Therefore, we fabricate a Au electrode using Au thin-film edges as a non-magnetic electrode and synthesize a BTBT-based chiral molecule, *N*-(3*S*)-3,7-dimethyloctyl[1]benzothieno[3,2-*b*]benzothiophene-2-carboxyamide (*S*-BTBT-CONHR), for the realization of nanoscale MR devices using chiral molecules. To observe the MR effect based on the CISS, we fabricate Au/*S*-BTBT-CONHR/$Ni_{78}Fe_{22}$ nanodevices using both Au and $Ni_{78}Fe_{22}$ thin-film edges, as shown in Fig. 1(a). As shown in Fig. 1(b), the highest occupied molecular orbital (HOMO) and the lowest unoccupied molecular orbital (LUMO) are calculated as −5.706 and −1.617 eV, respectively, which are sufficient for carrier injection from the Au or $Ni_{78}Fe_{22}$ electrode to the HOMO level. The mc-AFM studies reveal that *S*-BTBT-CONHR exhibits a high degree of spin selectivity. The MR effect has been successfully observed in the fabricated nanodevices under a low magnetic field at room temperature. These MR curves correspond to the magnetization curve of the $Ni_{78}Fe_{22}$ electrode, indicating that the fabricated nanodevices exhibit an MR effect based on the CISS effect originating from *S*-BTBT-CONHR. These results can lead to the development of nanoscale MR devices based on the CISS effect under low magnetic fields. However, despite the nanoscale junction area, the MR ratio is not as high as that predicted from the degree of spin selectivity by mc-AFM obtained in *S*-BTBT-CONHR thin films. A low MR ratio in nanodevices implies that there should be other causes for the low MR ratio, except for the leakage current through pinholes in chiral molecules. This study provides new insights into the CISS effect mechanism on devices.

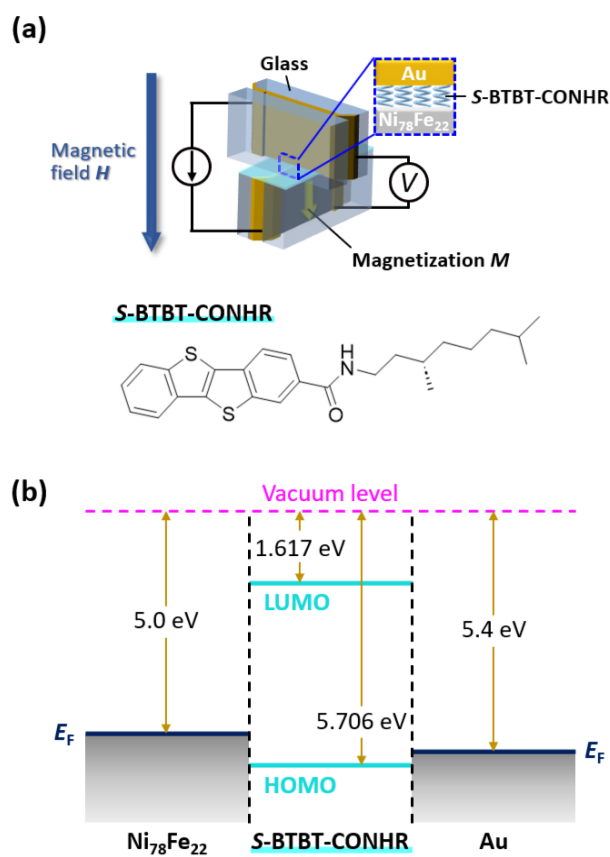

**Fig. 1** (a) Schematic of Au/*S*-BTBT-CONHR/Ni$_{78}$Fe$_{22}$ nanodevices. The nanodevice consists of the BTBT-based chiral molecule (*S*-BTBT-CONHR) sandwiched between Au and Ni$_{78}$Fe$_{22}$ thin films with their crossed edges. (b) Energy diagram of the nanodevices, showing the work functions of the Au and Ni$_{78}$Fe$_{22}$ electrodes[37,38] and the calculated HOMO–LUMO levels of *S*-BTBT-CONHR.

## Experimental section

### Preparation of *S*-BTBT-CONHR

Thionyl chloride (1.8 mL) and 3 drops of dry triethylamine were added to a solution of [1]benzothieno[3,2-*b*]benzothiophene-2-carboxylic acid[39] (1.00 g, 3.53 mmol) in dry toluene (50 mL). The reaction mixture was refluxed for 18 h, then cooled down to 25 °C. After the removal of thionyl chloride and toluene under reduced pressure, dry tetrahydrofuran (30 mL) was added. A solution of *S*-citronellamine[40,41] (1.38 g, 8.76 mmol) in dry tetrahydrofuran (20 mL) and dry triethylamine (1.0 mL, 7.2 mmol) was then added to the solution, which was stirred for 16 h at 50 °C. After cooling the reaction mixture to 25 °C, the white precipitate was removed by filtration. Methanol was added to the filtrate and the resulting white precipitate was filtered and washed with methanol. The crude product was recrystallized twice from tetrahydrofuran/acetonitrile (v/v = 1/4) to obtain white crystals (892 mg) in 60% yield. Elemental analysis: Calculated for $C_{25}H_{29}NOS_2$: C, 70.88; H, 6.96; N, 3.31. Found: C, 70.69; H, 6.80; N, 3.29. Additional experimental results for bulk *S*-BTBT-CONHR are shown in the Supplementary Information (SI, see Fig. S1–S7).

**Density functional theory (DFT) calculations**

The electronic structure of BTBT-CONHCH$_3$ was evaluated using DFT calculations prior to bulk characterization. For simplicity of calculation, BTBT-CONHCH$_3$ was used instead of *S*-BTBT-CONHR. The HOMO and LUMO levels were derived from the B3LYP/6-31g(d) basis set using a methylamide derivative because the alkyl chain length and chiral carbon derivatives are not expected to affect the electronic state of the BTBT π-electron core. The volume of *S*-BTBT-CONHR was estimated by DFT calculation using Gaussian 16 with B3LYP functional and 6-31G(d) basis sets. The computed molecular volume is defined here as the volume inside a contour of 0.001 electrons/Bohr$^3$ density.

**Device fabrication**

The fabrication procedure for the Au/*S*-BTBT-CONHR/Ni$_{78}$Fe$_{22}$ nanodevices consisted of sputtering, thermal pressing, mechanical cutting, polishing, and off-center spin-coating (OCSC), as shown in Fig. 2. Au and Ni$_{78}$Fe$_{22}$ thin films were deposited on polished low-softening-point (LSP) glass (Fig. 2(a)) using a DC magnetron sputtering system (Fig. 2(b)). A base pressure in the range of $3.0–6.9 \times 10^{-6}$ Pa was maintained before sputtering. Growth rates of the Au and Ni$_{78}$Fe$_{22}$ thin films were approximately 6 and 8 nm min$^{-1}$ at sputtering powers of 20 and 100 W, respectively, under a pressure of $6.0 \times 10^{-1}$ Pa with Ar gas of 15 sccm. Au thin films with a thickness of 14 nm were sputtered onto the chamfered edges of the deposited Ni$_{78}$Fe$_{22}$ and Au thin films (Fig. 2(c)). The chamfered edges are to provide access for electrical contacts. After that, glass/Au/glass and glass/Ni$_{78}$Fe$_{22}$/glass samples were fabricated by thermal pressing at pressures of 1.00 and 0.75 MPa, followed by cutting (Fig. 2(d), (e)). The thicknesses of the Au and Ni$_{78}$Fe$_{22}$ thin films were in the range of 42–50 nm. These Au and Ni$_{78}$Fe$_{22}$ thin-film edges can provide $42 \times 42–50 \times 50$ nm$^2$ nanojunctions. After polishing the cut surfaces of both glass/Au/glass and glass/Ni$_{78}$Fe$_{22}$/glass, *S*-BTBT-CONHR thin films were deposited on the polished surfaces by OCSC using dichloromethane/ethanol solution (v/v = 4/1) with a concentration of 6.0 mg mL$^{-1}$ (Fig. 2(f)). The OCSC process was performed at 500 rpm for 30 s with an acceleration of 50 rpm s$^{-1}$. Finally, Au/*S*-BTBT-CONHR/Ni$_{78}$Fe$_{22}$ nanodevices were fabricated by sandwiching the *S*-BTBT-CONHR thin films between glass/Au/glass and glass/Ni$_{78}$Fe$_{22}$/glass electrodes with the Au and Ni$_{78}$Fe$_{22}$ thin-film edges crossed, as shown in Fig. 2(g).

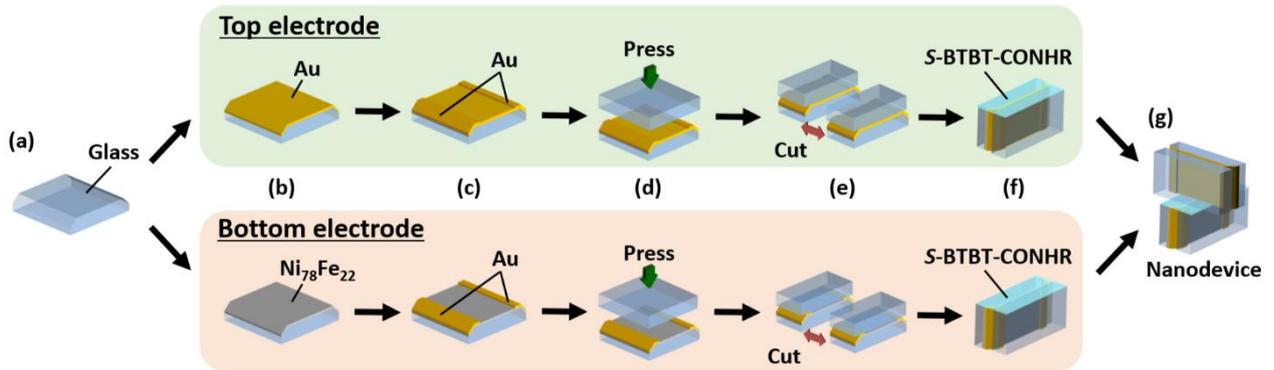

**Fig. 2** Fabrication procedure of Au/*S*-BTBT-CONHR/Ni$_{78}$Fe$_{22}$ nanodevices. (a) Preparation of LSP glasses upper (lower): (b) sputtering Au (Ni$_{78}$Fe$_{22}$) thin film onto the glass, (c) sputtering Au on the chamfered edges, (d) thermal pressing to fabricate glass/Au/glass (glass/Ni$_{78}$Fe$_{22}$/glass), (e) cutting glass/Au/glass (glass/Ni$_{78}$Fe$_{22}$/glass), (f) polishing the cut surface of glass/Au/glass (glass/Ni$_{78}$Fe$_{22}$/glass) and spin-coating *S*-BTBT-CONHR thin film on the polished surface. (g) Fabricating Au/*S*-BTBT-CONHR/Ni$_{78}$Fe$_{22}$ nanodevices by sandwiching the *S*-BTBT-CONHR thin films between the Au and Ni$_{78}$Fe$_{22}$ thin-film edges.

**Sample preparation for mc-AFM measurements**

*S*-BTBT-CONHR thin films were spin-coated on the freshly cleaved highly oriented pyrolytic graphite (HOPG) substrates using dichloromethane/ethanol solutions (v/v = 4/1) with concentrations of 0.5 and 6.0 mg mL$^{-1}$. The spin-coating process was performed at 500 rpm for 5 s with an acceleration of 100 rpm s$^{-1}$, followed at 3000 rpm for 30 s with an acceleration of 500 rpm s$^{-1}$.

**Measurement setup**

The interfacial features of the glass/Au/glass samples were examined by transmission electron microscopy (TEM) and energy-dispersive X-ray spectroscopy (EDS) (TECNAI Osiris, FEI). Cross-sectional TEM specimens were prepared using the focused ion beam (FIB) technique. The surface morphologies and roughness of the polished glass/Au/glass substrates coated with and without *S*-BTBT-CONHR were analyzed using AFM (Nanocute, SII Nano Technology Inc.). The electrical properties of the Au thin-film edges coated with and without *S*-BTBT-CONHR were evaluated using a c-AFM system equipped with a Rh-coated cantilever (Si–DF3–R, SII Nano Technology Inc.). In the mc-AFM studies, *I–V* curves were measured using a c-AFM system with a CoPtCr-coated cantilever (SI–MF20, SII Nano Technology Inc.). Before *I–V* measurements, we applied a magnetic field of approximately ±300 Oe to the magnetic cantilever. The magnetization was maintained along the up or down direction during the measurements without an external magnetic field. Forty *I–V* curves were averaged for each magnetization state of the cantilever. For each *I–V* measurement, the cantilever was placed in a new position. The magnetic properties of the $Ni_{78}Fe_{22}$ thin films were evaluated using magneto-optical Kerr effect spectroscopy (MOKE; BH-PI920-HU, NEOARK). The MR effect of devices was investigated by a four-probe method under a magnetic field up to ±190 Oe at room temperature.

## Results and discussion

### Fabrication of the Au electrodes

The cross-sectional TEM image of the glass/Au/glass specimen exhibits a Au thin film with smooth and well-defined interfaces, as shown in Fig. 3(a). After the 50-nm-thick Au thin film was thermally pressed under a pressure of 1.00 MPa, the pressed Au thin film has a thickness of 48 nm obtained from Fig. 3(a). Fig. 3(b) and (c) show the high-angle annular dark field (HAADF) and EDS mapping of the Au images of the same sample. The glass/Au/glass specimen exhibits smooth and clear interfaces with no diffusion of Au atoms into the LSP glass. The results reveal that the glass/Au/glass sample can be successfully fabricated by thermal pressing at a pressure of 1.00 MPa. Fig. 3(d) shows an AFM image of the polished cross-sectional surface of the glass/Au/glass specimen. The roughness of the polished surface is 0.63 nm measured over a scanning area of 10 × 20 μm². Fig. 3(e) shows a c-AFM image of the same scanning area as shown in Fig. 3(d), exhibiting uniform electrical conduction along the Au thin-film edges. These microscopic studies indicate that the fabricated glass/Au/glass specimen can be used as an electrode in the proposed nanodevices. To achieve both low roughness and uniform electrical conduction in the cross-sectional surface of the glass/Au/glass specimen, the polishing time for this specimen is optimized and changed to 1.2 times longer than that of the glass/$Ni_{78}Fe_{22}$/glass specimen in the finishing chemical mechanical polishing process using colloidal $SiO_2$ slurries. Because we have already established a fabrication method for glass/Ni$_{78}$Fe$_{22}$/glass, we can fabricate glass/Ni$_{78}$Fe$_{22}$/glass with a low polished surface roughness and uniform electrical conduction along the Ni$_{78}$Fe$_{22}$ edges, as presented in our previous study.[9] The results of the structural and electrical analyses of the glass/Ni$_{78}$Fe$_{22}$/glass specimen are shown in Fig. S8.

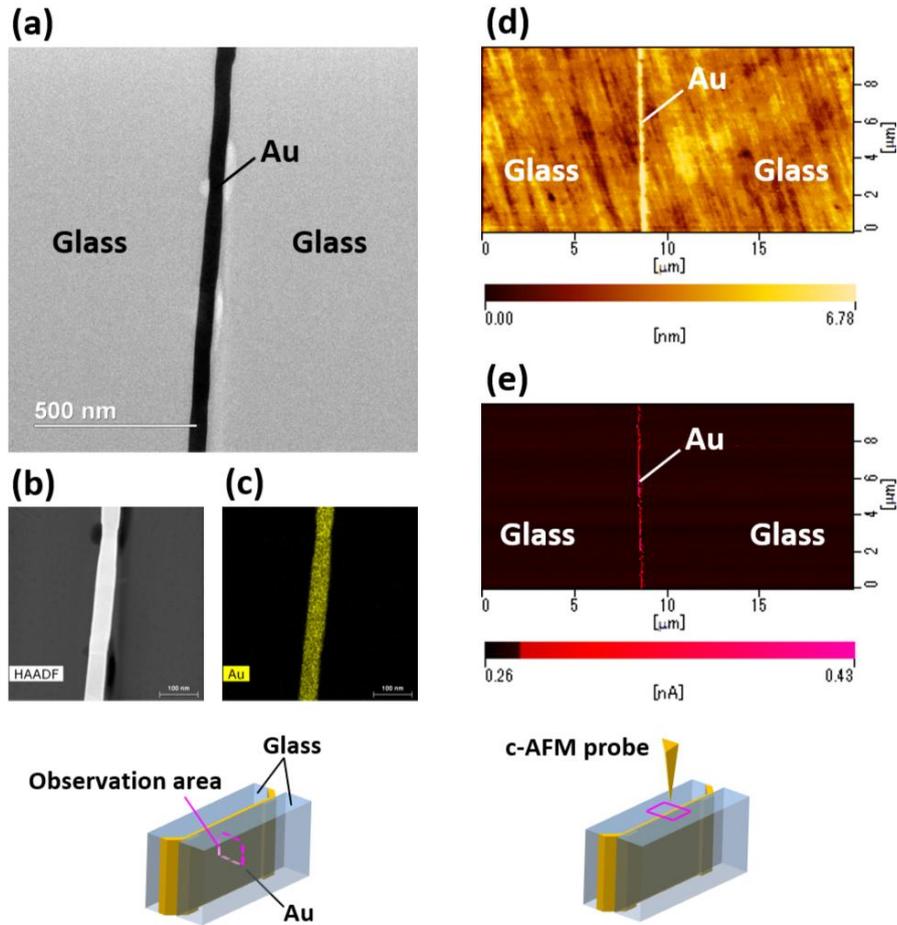

**Fig. 3** (**a**) Cross-sectional TEM, (**b**) HAADF, and (**c**) EDS mapping of Au images of the 48-nm-thick Au thin film sandwiched between the two LSP glasses. (**d**) AFM and (**e**) c-AFM images of the polished surface of the glass/Au/glass samples.

### *S*-BTBT-CONHR thin films on the Au thin-film edges

Fig. 4(a) shows the cross-sectional TEM image of the 1-nm-thick *S*-BTBT-CONHR thin film on the polished surface of the glass/Au/glass sample. An *S*-BTBT-CONHR film with a smooth and clear interface is successfully deposited on the Au thin-film edges. Fig. 4(b) and (c) show AFM and c-AFM images of the surface of the *S*-BTBT-CONHR thin film on the polished surface of glass/Au/glass samples, respectively. We have also achieved both low roughness (0.76 nm) and uniform electrical conduction along the Au edges on the surface of the glass/Au/glass specimen coated with the *S*-BTBT-CONHR thin film. Therefore, using glass/Au/glass and glass/Ni$_{78}$Fe$_{22}$/glass samples coated with *S*-BTBT-CONHR thin film, we can fabricate nanodevices consisting of an *S*-BTBT-CONHR layer sandwiched between the Au and Ni$_{78}$Fe$_{22}$ edges. In addition, the DFT calculation result reveals that the volume of *S*-BTBT-CONHR is 337.4 cm$^3$/mol, which corresponds to approximately 560 Å$^3$ per molecule. Based on the *S*-BTBT-CONHR film thickness (totally ~2 nm), junction area (48 × 49 nm$^2$) and the volume of *S*-BTBT-CONHR (560 Å$^3$), the number of the accommodated molecules can be estimated to be approximately 8400 in our nanojunctions.

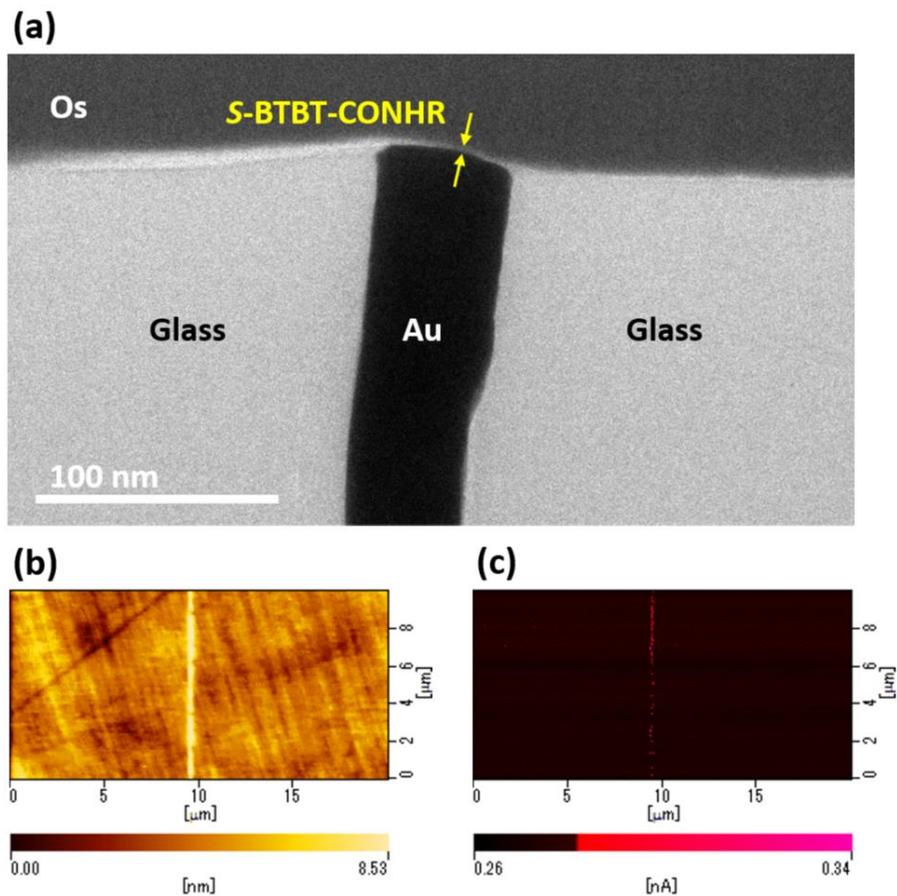

**Fig. 4** (**a**) Cross-sectional TEM image obtained for the 1-nm-thick *S*-BTBT-CONHR thin film on the polished surface of the glass/Au/glass sample. (**b**) AFM and (**c**) c-AFM images of the surface of the *S*-BTBT-CONHR thin film on the polished surface of the glass/Au/glass sample.

**mc-AFM studies of the *S*-BTBT-CONHR thin films**

Fig. 5(a) shows a schematic of the setup used for the mc-AFM measurements. Using this setup, *I–V* characteristics are obtained for the *S*-BTBT-CONHR thin film, as shown in Fig. 5(b) and (c). 6 mg/ml solution provided a thicker film than that formed by 0.5 mg/ml solution. The surface morphology and thickness estimation of the *S*-BTBT-CONHR thin film on HOPG are shown in Fig. S9 and S10. The raw data of the *I–V* curves and the averaged *I–V* plots with the error bar which consists of 1/10 plots from the original averaged data for each film are shown in Fig. S11 and S12. In the *I–V* curves shown in Fig. 5(b) and (c), $|I_{down}|$ is larger than $|I_{up}|$ at each bias voltage. These results indicate that *S*-BTBT-CONHR exhibits the CISS effect and selects the same-direction spin independent of both the magnitude and sign of the bias voltage. This independence is consistent with mc-AFM results for chiral molecules reported in previous studies.[18–30] From Fig. 5(b) and (c), the degree of spin selectivity, $SS_{CISS}$, is calculated by using $(I_{down} − I_{up})/(I_{down} + I_{up})$ at each bias voltage, and Fig. 5(d) shows the relationships between $SS_{CISS}$ and bias voltage. The average $SS_{CISS}$ is as high as 0.87 (0.72) in the *S*-BTBT-CONHR thin film spin-coated using the solution with a concentration of 0.5 (6.0) mg mL$^{-1}$. The reason why there is no significant difference in $SS_{CISS}$ is that our chiral films consist entirely of organic molecules. In the recent study, the 180-μm-thick chiral film provided a high $SS_{CISS}$.[28] This chiral film consisted of a combination of inorganic and organic materials, which cannot be easily compared with our results using only organic molecules. In addition, the same group and other groups investigated the CISS effect by mc-AFM in the monolayer of chiral molecules[19,23,30] In our case, the chiral film spin-coated by a solution of 0.5 mg mL$^{-1}$ is considered to be almost monolayer. On the contrary, in the thicker film spin-coated by a solution of 6.0 mg mL$^{-1}$, the electrical conduction may be not tunneling but hopping. There is a high possibility that hopping conduction can generate spin scattering, reducing $SS_{CISS}$. This explanation is consistent with the previous study using the 180-μm-thick chiral film consisting of the inorganic and organic materials because there are many times of tunnel conduction from a metal to metal through a chiral molecule. Moreover, the study reported that MR exhibited a weak thickness dependence in the organic/inorganic hybrid perovskite systems because the slight thickness dependence originates from the two competing processes; spin-selective increase due to the increase in the number of chiral barrier layers and spin relaxation due to spin scattering.[27] Therefore, our explanation is also consistent with the study. In some cases, the 20–120-nm-thick chiral films (not contain inorganic materials) provided a high spin selectivity which may originate from the high conductivity of the molecules.[24,29] However, these behaviors can be changed by various factors, such as the size, length, density of electrons and crystallinity of chiral molecules, the work function of electrode materials and fabrication method. These should continue to be investigated to clarify the unified mechanism of the CISS effect. As shown in Fig. 5(d), $SS_{CISS}$ increases (decreases) when the bias voltage increases in the *S*-BTBT-CONHR thin film spin-coated using solutions with a concentration of 0.5 (6.0) mg mL$^{-1}$. In previous mc-AFM studies, $SS_{CISS}$ was enhanced using bias voltage in a monolayer of chiral molecules,[23] which is consistent with the bias voltage dependence observed in the *S*-BTBT-CONHR thin film spin-coated using a solution with a concentration of 0.5 mg mL$^{-1}$. Conversely, another study reported that $SS_{CISS}$ was reduced by bias voltage in the amorphous thin film of the chiral molecule with a thickness of approximately 22 nm,[24] which is consistent with the *S*-BTBT-CONHR thin film spin-coated using a solution with a concentration of 6.0 mg mL$^{-1}$. Thus, our results indicate that the bias voltage dependence of the CISS effect is influenced by the thickness of the molecular films and/or the molecular orientation. Here, we note that although many mc-AFM measurements have been performed in previous studies, unified rules have not been established, and the results cannot be easily compared because of different setups and the ambiguous definition of up or down magnetization in some cases. Our mc-AFM studies clarify the different behaviours of the CISS effect between the two thin films, as shown in Fig. 5(b)–(d), because a unified setup was used. Thus, to clarify the CISS effect mechanism, a key issue is that the same research group should perform mc-AFM measurements on various chiral molecules using a unified setup. In addition, it is important to report these results with an exact explanation of the setup used, such as the configurations of the up or down magnetizations.

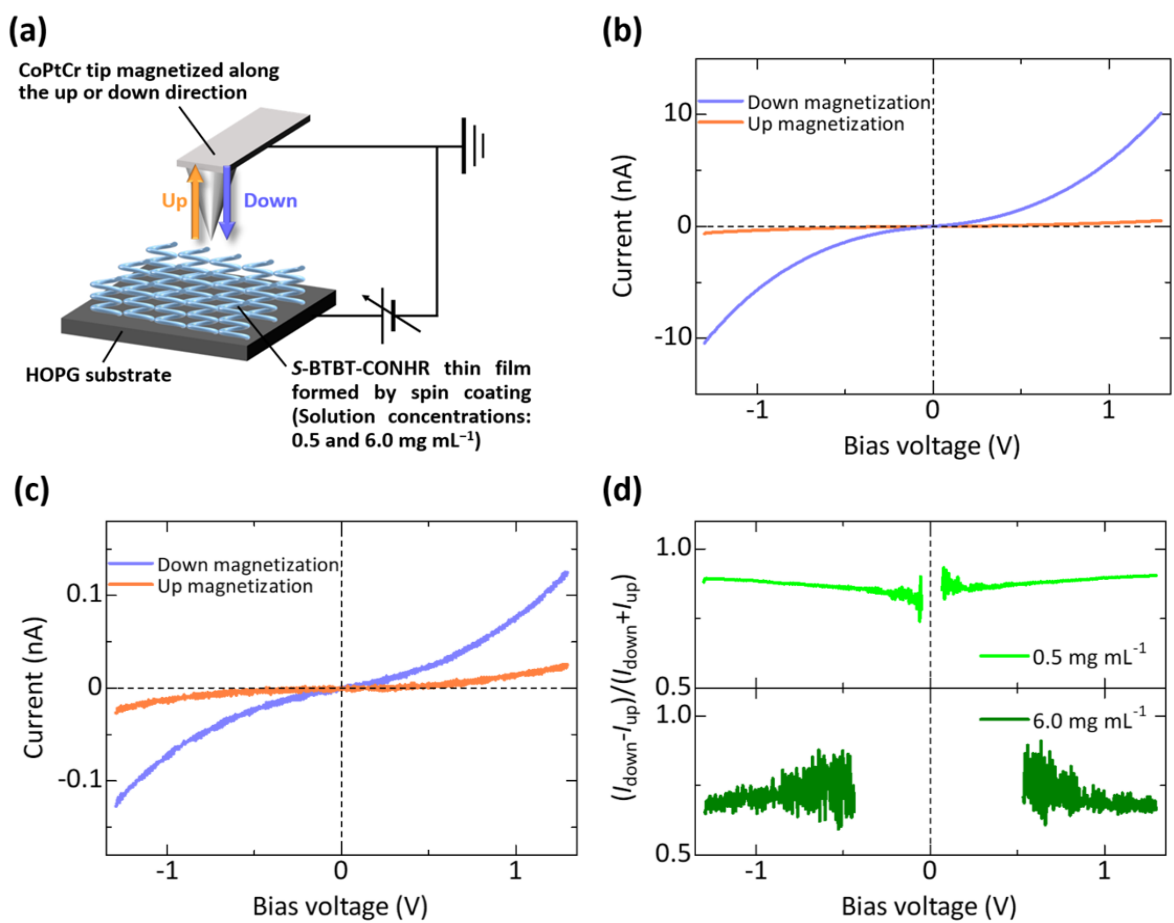

**Fig. 5** (**a**) Schematic of the setup for mc-AFM measurements for *I*–*V* characteristics in the *S*-BTBT-CONHR thin film spin-coated using solutions with concentrations of (**b**) 0.5 and (**c**) 6.0 mg mL$^{-1}$, respectively. (**d**) Bias voltage dependence of the degree of spin selectivity, SS$_{CISS}$, calculated by using ($I_{down} - I_{up}$)/($I_{down} + I_{up}$).

**MR effect in Au/*S*-BTBT-CONHR/Ni$_{78}$Fe$_{22}$ nanodevices**

As shown in Fig. 6(a), the magnetization curve of the Ni$_{78}$Fe$_{22}$ thin film exhibits a coercivity of 39 Oe, which is higher than that of 4 Oe before thermal pressing (Fig. S13). This increase in coercivity can be explained by the random anisotropy model, as discussed in our previous studies.[9,42] Using the described methods, we fabricate Au/*S*-BTBT-CONHR/Ni$_{78}$Fe$_{22}$ nanodevices using a Ni$_{78}$Fe$_{22}$ electrode with a coercivity of 39 Oe (Fig. 6(a)) and investigate the MR effect at room temperature. We also fabricate Au/Ni$_{78}$Fe$_{22}$ nanodevices without a chiral molecular layer, in which no MR effect is observed (Fig. S14). In the fabricated Au/*S*-BTBT-CONHR/Ni$_{78}$Fe$_{22}$ nanodevices, the MR effect is successfully observed, as shown in Fig. 6(b), which is colored red and blue according to the direction of the sweeping field. These MR curves correspond to the magnetization states of the $Ni_{78}Fe_{22}$ electrode, indicating that a clear MR effect based on the CISS effect is successfully observed in the nanodevices using chiral molecules. In addition, we can observe the CISS-based MR effect under the lowest magnetic field ever reported for OSVs using chiral molecules.[20,22,23,25–29,34–36] Generally, OSVs have layered structures, and a Ni thin film is used as the magnetic layer in CISS-based OSVs. A high magnetic field of approximately 5 kOe is required to magnetize the Ni thin film along the perpendicular direction. Recently, instead of Ni thin films, Pt/Co bilayers have been used as the magnetic layers in CISS-based OSVs because they exhibit perpendicular magnetic anisotropy (PMA) and low coercivity.[36] Using a magnetic electrode with PMA, these OSVs can exhibit MR curves with magnetic hysteresis corresponding to the magnetization states of the magnetic electrode. Therefore, the CISS-based MR effect was observed with magnetic hysteresis in OSVs utilizing a Pt/Co bilayer under a low magnetic field of approximately 200 Oe because of the low coercivity.[36] Compared with the method using a Pt/Co bilayer, we can realize OSVs with PMA and low coercivity more easily because of our unique device structure utilizing a magnetic thin film with in-plane magnetic anisotropy, which is perpendicular to the molecular layer. In other words, we can fabricate CISS-based OSVs with PMA and low coercivity using various magnetic materials, such as FeCo and Co, in addition to Ni, which is often used as a magnetic layer in CISS-based OSVs. FeCo and Co can improve the MR ratio because of their high spin polarizations $P$. In these devices with various magnetic materials, an MR effect with magnetic hysteresis is expected to occur under low magnetic fields. Thus, our fabrication method is useful for investigating the relationships between many types of magnetic materials and the CISS-based MR effects in devices.

In addition, the MR curves exhibit the same polarity at both positive and negative bias voltages (Fig. S15). This indicates that the same-direction spin is selected independently of the spin-injection direction in the devices, which is consistent with previous results[18–30] as well as our results obtained by mc-AFM studies (Fig. 5). However, the MR ratio is less than 0.1% in the nanoscale devices fabricated in this study, although previous studies have suggested that the low MR ratio is caused by pinholes in the chiral molecular layer of the devices with milli- or microscale junction areas. The low MR ratio in nanodevices implies that there are causes other than the leakage current through the pinholes in the chiral molecules. Here, it is difficult to ensure no atomic level pinhole in our nanodevices. The atomic level pinhole can also be detected in the mc-AFM study. Thus, the presence of atomic level pinholes cannot be perfectly avoided in any devices and/or methods. From this viewpoint, we compared the mc-AFM results with the MR results in the nanodevices. As described earlier, it should be noted that we could not observe any CISS-based MR effect in the Au/Ni$_{78}$Fe$_{22}$ device (Fig. S14), in which Au is directly contacted with Ni$_{78}$Fe$_{22}$. This result indicates that our nanodevices can successfully sandwich the chiral films, exhibiting the CISS-based MR effect. Although there might be atomic level pinholes in our nanodevices, the number of pinholes should be less than that in the microscale devices reported from the previous studies.[20,22,23,25–29,34–36] One of the alternative possibilities for the low MR in the devices is an orientation of chiral molecules. In our study, the surface roughness of chiral films on the HOPG substrate is lower than that on the electrodes of nanodevices. Considering the value of the surface roughness, the orientation of the chiral molecules can be different from each other. Another possibility is a magnitude of the bias voltage. In the mc-AFM study, the applied voltage is over 1 V, whereas it is a few millivolts in our nanodevices because of its low resistance. Although the mechanism of low resistance has not been clarified, the low resistance has been observed reproducibly. If we can apply the high voltage in the nanodevices, high MR might be observed. Considering the previous studies, their devices sandwich the inorganic insulator such as Al$_2$O$_3$ and MgO, while the mc-AFM setup does not contain inorganic insulators.[20,22,23,25,26,28,29] These results imply that the inorganic insulator may reduce MR because of increasing spin scattering and/or the difference in the interfacial state of the chiral molecular layer. On the other hand, in the previous studies, although the leakage current through the pinholes could be suppressed using the inorganic insulators, a low MR ratio was observed. This discussion also indicates that there is another cause for low MR instead of the pinholes in chiral films. In the research field of the CISS effect, it is important that a high MR is achieved in the devices for practical applications in the future. The CISS effect in the devices should continue to be investigated and discussed, and our study can contribute to its development.

*R*-BTBT-CONHR-based experiments can emphasize that our results (both mc-AFM studies and MR in the devices) originate from the CISS effect. On the other hand, we provide the references for the experiments using a molecule with only one chirality.[19,43,44] Here, we fabricated the Au/*racemic*-BTBT-CONHR/Ni$_{78}$Fe$_{22}$ device, in which the one-side electrode is coated with *racemic*-BTBT-CONHR because we have focused on the possibility that the nanoscale junction region could sandwich the chiral layer enriched in the *R*- or *S*-derivative due to the formation of separate crystalline domains of the *R*- and *S*-derivatives in the *racemic*-BTBT-CONHR thin film. In this device, as shown in Fig. S16, the inverse MR effect is observed with respect to the MR polarity of the *S*-BTBT-CONHR device. This can be attributed to the formation of separate crystalline domains of the *R*- and *S*-derivatives within the molecular films composed of the *racemic* mixture, which leads to the presence of a nanoscale region enriched in the *R*-derivative at the junction. Therefore, these results indicate that the observed MR (Fig. 6(b), Fig. S16 and S17) originate from the CISS effect of chiral BTBT.

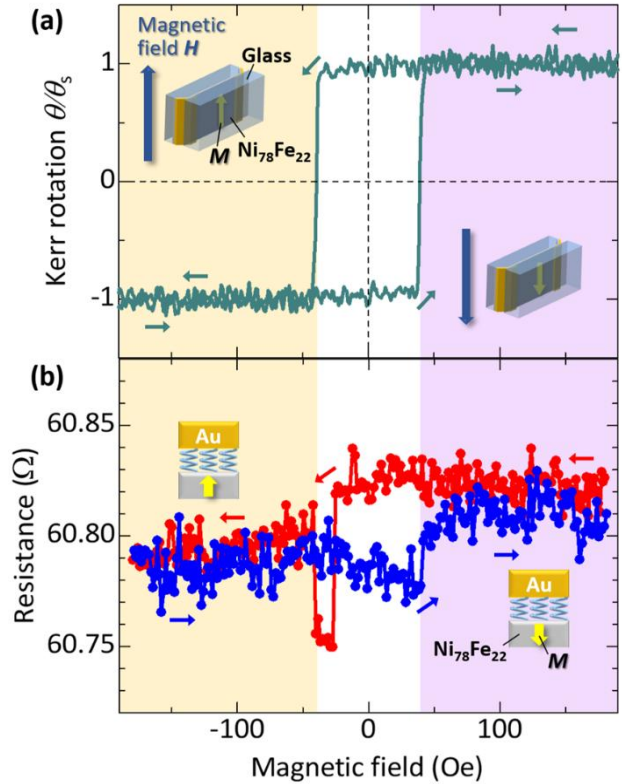

**Fig. 6** (**a**) Magnetization curve of the $Ni_{78}Fe_{22}$ electrode with a coercivity of 39 Oe. (**b**) MR curves in Au/*S*-BTBT-CONHR/$Ni_{78}Fe_{22}$ nanodevices. The blue (red) plots represent the results obtained under the forward (reverse) sweeping magnetic field.

## Conclusions

To observe the MR effect based on the CISS, we aim to realize CISS-based nanodevices using both Au and $Ni_{78}Fe_{22}$ thin-film edges. As the Au electrode in the proposed nanodevices, glass/Au/glass can be successfully fabricated using a thermal pressing technique, and the Au thin-film edges show uniform electrical conduction by polishing under optimized conditions. In addition, we have synthesized *S*-BTBT-CONHR for the nanodevices, which can show a high degree of spin selectivity, as revealed by mc-AFM studies. Using these electrodes and chiral molecules, we have successfully fabricated Au/*S*-BTBT-CONHR/Ni$_{78}$Fe$_{22}$ nanodevices. In this chiral-molecule-based MR device, the MR effect can be observed under a low magnetic field at room temperature. These MR curves correspond to the magnetization states of the Ni$_{78}$Fe$_{22}$ electrode. This indicates that the MR effect based on the CISS effect is successfully observed in nanodevices using BTBT-based chiral molecules. This study can lead to the development of CISS-based MR devices under low magnetic fields and provide new insights into the CISS effect mechanism on devices.

## Author contributions

M. M. and H. K. conceived and designed the experiments. M. M., R. M., T. U. and K. T. fabricated the electrodes. K. S. and T. A. synthesized the chiral molecules and evaluated their bulk physical properties. T. Y., M. M. and R. M. contributed to solution preparation and deposition of the molecular layers. M. M. and K. K. performed the mc-AFM measurements and analyses of their results. M. M., K. T. and T. U. analyzed the structural and electrical properties of the fabricated electrodes. K. T. measured the magnetic properties of the Ni$_{78}$Fe$_{22}$. M. M. performed the device fabrication, MR measurements, and analyses of the results. M. M. and T. A. wrote the original draft. H. K., T. Y., and the other authors contributed to the review and editing of the manuscript. H. K. and T. A. supervised the study. All authors contributed to data interpretation.

## Conflicts of interest

There are no conflicts to declare.

## Data availability statements

The datasets used and/or analyzed in the current study are available from the corresponding author upon reasonable request.

## Acknowledgements

The authors would like to express their sincere appreciation to the technical staff members of the Central Service Facilities for Research of Keio University for their assistance with FIB processing, TEM observations, and EDS analyses. We also thank Shunki Kashii as well as Shun Dekura of Tohoku University for the DFT calculations and Masayuki Suda of Kyoto University for fruitful discussions. This research was supported by a Grant-in-Aid for Scientific Research (B) (No. 24K00948) and a Grant-in-Aid for the Japan Society for the Promotion of Science (JSPS) Fellows (No. 24KJ1953), funded by the JSPS, Japan Science and Technology Agency Support for Pioneering Research Initiated by the Next Generation (JST SPRING; No. JPMJSP2123), and Center for Spintronics Research Network (CSRN) at Keio University.

# Magnetoresistance effect based on spin-selective transport in nanodevices using chiral molecules

# Supplementary Information

Mizuki Matsuzaka,[a] Kotaro Kashima,[a] Koki Terai,[a] Takumi Ueda,[a] Ryunosuke Miyamoto,[a] Takashi Yamamoto,[a] Kohei Sambe,[b] Tomoyuki Akutagawa[b] and Hideo Kaiju[*ac]

[a]*Faculty of Science and Technology, Keio University, Yokohama, Kanagawa 223-8522, Japan*
[b]*Institute of Multidisciplinary Research for Advanced Materials, Tohoku University, Sendai, Miyagi 980-8577, Japan*
[c]*Center for Spintronics Research Network, Keio University, Yokohama, Kanagawa 223-8522, Japan*

[*]Correspondence and requests for materials should be addressed to H. K. (email: kaiju@appi.keio.ac.jp).

In this Supplementary Information (SI) section, we show some experimental results that were not included in the main text. At first, we show the bulk physical properties of *S*-BTBT-CONHR. In addition, we present the structural and electrical properties of glass/Ni$_{78}$Fe$_{22}$/glass samples coated without and with 1-nm-thick *S*-BTBT-CONHR thin films. We also show surface morphologies and thickness estimation of *S*-BTBT-CONHR thin films on HOPG substrates for mc-AFM measurements and the raw data of *I*–*V* curves measured by mc-AFM. Moreover, we show magnetization curves measured in the Ni$_{78}$Fe$_{22}$ thin films before thermal pressing. Finally, we show the additional experimental results of the MR effect observed in the fabricated nanodevices.

**Bulk physical properties of *S*-BTBT-CONHR.** *S*-BTBT-CONHR was synthesized in 60% yield by the reaction of [1]benzothieno[3,2-*b*]benzothiophene-2-carboxylic acid chloride with *S*-citronellamine. Thermogravimetry (TG) analysis and differential scanning calorimetry (DSC) were conducted using a Rigaku Thermo plus TG8120 thermal analysis station and a Mettler DSC3-T unit with an Al$_2$O$_3$ reference at a heating and cooling rate of 5 K min$^{-1}$ under N$_2$-flow. *S*-BTBT-CONHR shows weight loss from 475 K in TG measurements, and high thermal stability was confirmed (Fig. S1). In DSC measurements, the as-grown crystals exhibited three phase transitions during heating at 422.0, 432.8, and 442.6 K with transition entropy changes Δ*S* = −21.7, −27.6, and −52.43 J mol$^{-1}$ K$^{-1}$, respectively (Fig. S2). The phase transitions at 422.0 and 432.8 K correspond to solid-solid transitions of S1-S2 and S2-S3, while that at 442.6 K corresponds to the S3-liquid (L) phase transition, which is the melting point. In subsequent thermal cycles, it reversibly showed S2-S3 (Δ*S* = −28.6 J mol$^{-1}$ K$^{-1}$) at 437.9 K and S3-L phase (Δ*S* = −33.3 J mol$^{-1}$ K$^{-1}$) transitions at 443.0 K.

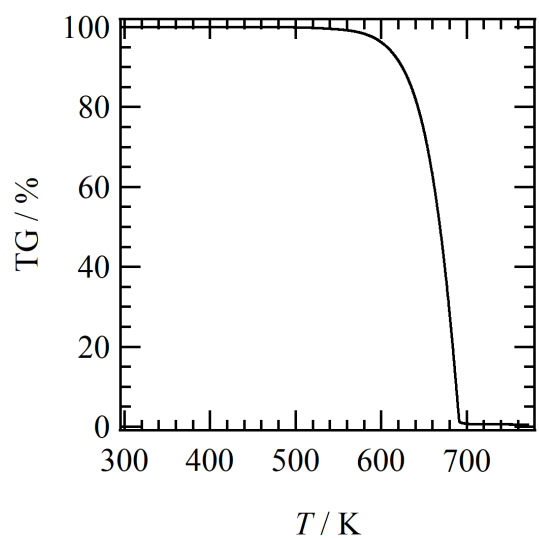

**Fig. S1** TG chart of *S*-BTBT-CONHR.

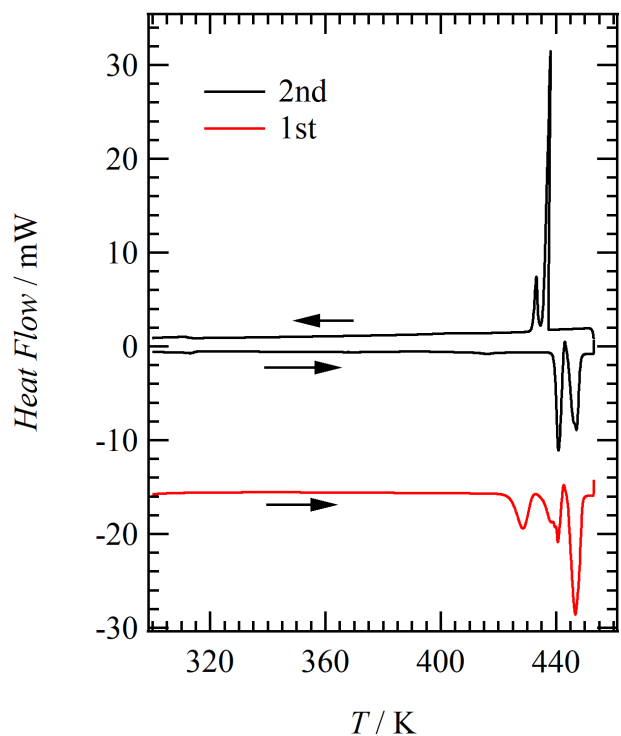

**Fig. S2** DSC charts of *S*-BTBT-CONHR in the 1st and 2nd thermal cycles.

Temperature-dependent powder X-ray diffraction (PXRD) patterns were obtained using a Rigaku Rint-Ultima diffractometer with Cu Kα radiation at $\lambda = 1.54185$ Å within the temperature range of 298–498 K. The PXRD pattern of *S*-BTBT-CONHR crystals showed sharp Bragg reflections at 300 K, indicating high crystallinity (Fig. S3). Peaks corresponding to an interlayer spacing of $d$ = 4.6 nm were observed in the low-angle $2\theta$ region along with their higher-order reflection patterns, confirming the formation of a layered structure. From the optimized molecular length is 2.2 nm, a layered structure with a bilayer periodicity was consistent with the interlayer spacing. After melting once, the number of reflection peaks at 300 K decreased, while upon heating, the number of reflection peaks clearly increased from around 430 K, showing the peaks corresponding to an interlayer distance of $d$ = 4.2 nm at 442 K. Microscopic observations were made using a Nikon ECLIPSE LV100ND, and stereomicroscopic and polarized light micrographs were taken using a Canon EOS kiss x5. No liquid crystal phase formation was confirmed by optical microscope (OM) images, while polarized OM (POM) images showed changes in texture when the temperature was raised from 20 °C to 100 °C (Fig. S4). Additionally, an AFM (Hitachi AFM5000) was used to observe the surface morphologies. AFM images at 30 °C and 80 °C revealed clear changes in domain structure (Fig. S5), with high-temperature treatment resulting in uniform changes to the domain structure on the substrate surface. Furthermore, the PXRD pattern of the thin film annealed at 425 K clearly showed reflection peaks reflecting the layered structure, exhibiting higher crystallinity (Fig. S6).

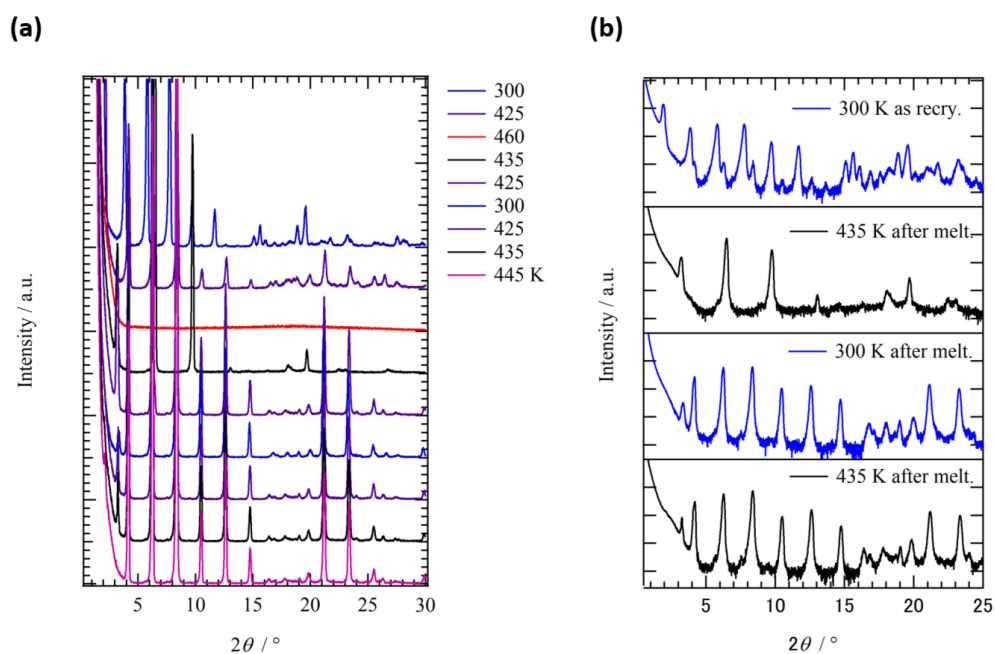

**Fig. S3** Temperature-dependent PXRD patterns of *S*-BTBT-CONHR. (a) Overall PXRD patterns and (b) the representative ones.

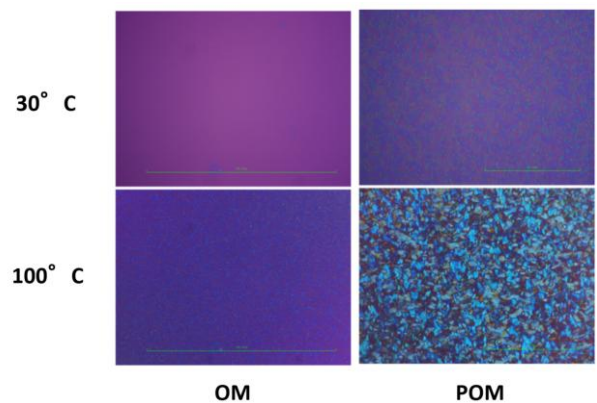

**Fig. S4** Temperature-dependent OM and POM images of *S*-BTBT-CONHR.

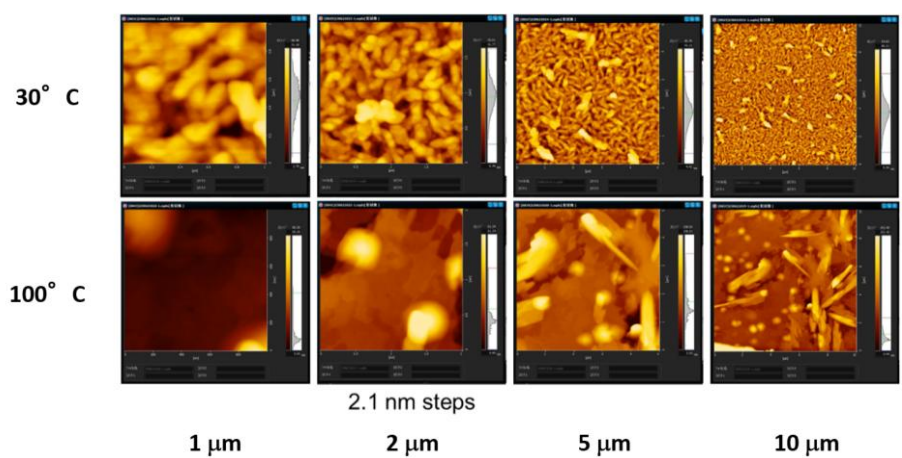

**Fig. S5** AFM images of vacuum deposited thin films of *S*-BTBT-CONHR at 30 °C (upper) and 100 °C (lower).

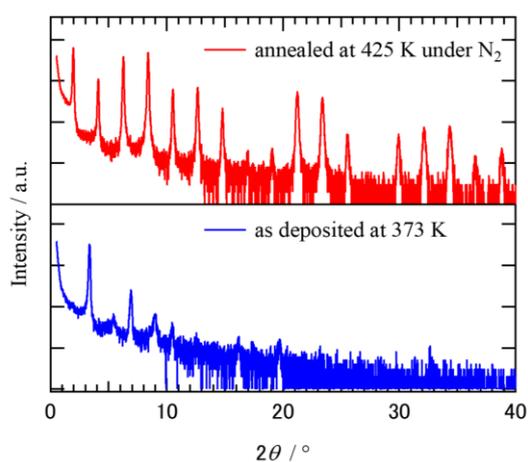

**Fig. S6** Out of plane PXRD patterns of vacuum deposited thin films of *S*-BTBT-CONHR as deposited (lower) and annealed at 425 K under $N_2$ (upper).

The temperature- and frequency-dependence of the dielectric constants of *S*-BTBT-CONHR showed that the real part ($\varepsilon_1$) monotonically decreased with increasing temperature and then increased depending on frequency from around 365 K (Fig. S7). It was confirmed that both the real and imaginary ($\varepsilon_2$) parts of the dielectric constants diverged around 440 K, where a peak was observed in the heating process of the DSC chart, indicating the presence of thermal motion of the amide group or chiral chain in bulk solid.

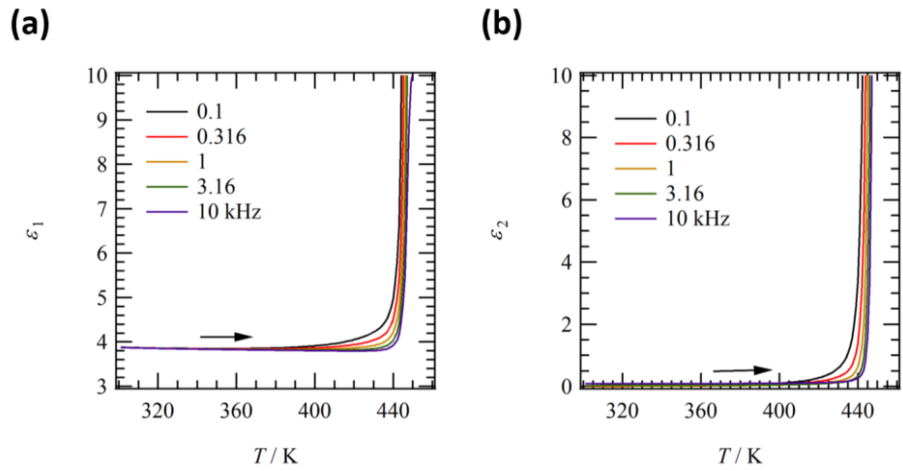

**Fig. S7** Temperature- and frequency-dependent (a) real part and (b) imaginary part dielectric constants of *S*-BTBT-CONHR.

**Structural and electrical properties of glass/Ni$_{78}$Fe$_{22}$/glass.** Fig. S8(a) shows the cross-sectional scanning electron microscopy (SEM; Inspect F50, FEI) image of the glass/Ni$_{78}$Fe$_{22}$/glass samples, exhibiting smooth and clear interfaces between glasses and Ni$_{78}$Fe$_{22}$ thin film. Prior to SEM imaging, a 4–5-nm-thick Os thin film was deposited on the polished surface of the glass/Ni$_{78}$Fe$_{22}$/glass samples by an osmium coater (HPC-20, Vacuum Device Inc.). From Fig. S8(a), the thickness of Ni$_{78}$Fe$_{22}$ thin film is found to be approximately 49 nm. This means that after the 55-nm-thick Ni$_{78}$Fe$_{22}$ thin film was thermally pressed under a pressure of 0.75 MPa, the pressed Ni$_{78}$Fe$_{22}$ thin film has a thickness of 49 nm. Here, ion-beam or DC magnetron sputtering systems were used in our previous study[1] or in this study, respectively, to deposit the Ni$_{78}$Fe$_{22}$ thin films onto the LSP glasses. This means that we can successfully fabricate the glass/Ni$_{78}$Fe$_{22}$/glass samples by using ion-beam or DC magnetron sputtering systems. Fig. S8(b)–(e) show the AFM and c-AFM images of the polished glass/Ni$_{78}$Fe$_{22}$/glass surfaces covered without and with 1-nm-thick *S*-BTBT-CONHR thin films. The obtained surface roughness is 0.85 (0.83) nm for the sample without (with) *S*-BTBT-CONHR thin films. The c-AFM images exhibit a uniform electrical conduction along the Ni$_{78}$Fe$_{22}$ edges coated without and with *S*-BTBT-CONHR thin films.

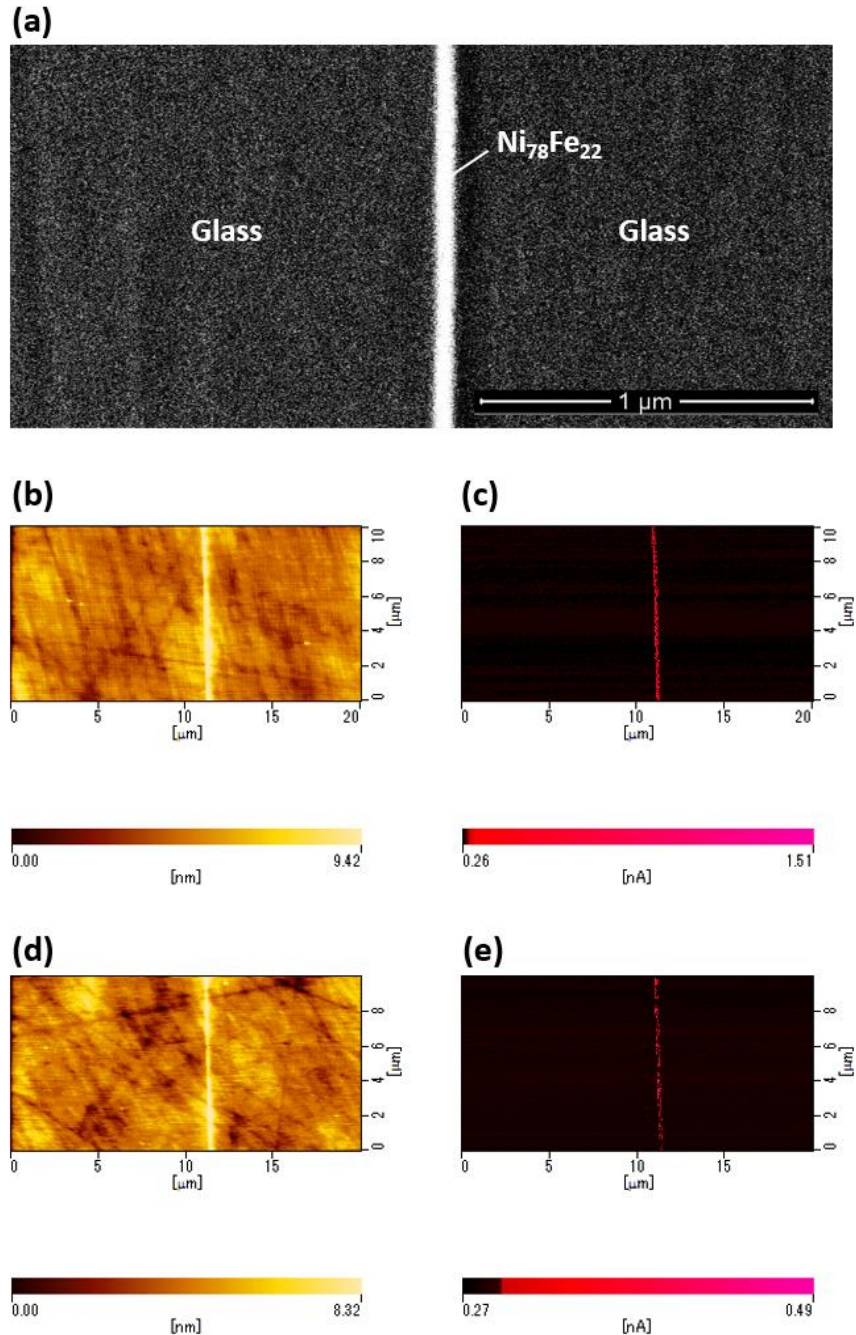

**Fig. S8** (a) Cross-sectional SEM image of the polished glass/Ni$_{78}$Fe$_{22}$/glass samples. AFM and c-AFM images of the polished glass/Ni$_{78}$Fe$_{22}$/glass samples coated (b, c) without and (d, e) with 1-nm-thick *S*-BTBT-CONHR thin films.

**Surface morphologies of the *S*-BTBT-CONHR thin films on HOPG substrates.** Prior to mc-AFM measurements, we evaluated the surface roughness of the *S*-BTBT-CONHR thin films on HOPG substrates. In this evaluation, we selected a dynamic force mode in AFM with a Si cantilever (Si–DF20, SII Nano Technology Inc.). Fig. S9 shows the AFM images of the *S*-BTBT-CONHR thin films on HOPG substrates spin-coated from the solutions with concentrations of 0.5 and 6.0 mg mL$^{-1}$. These images indicate that we can successfully fabricate the smooth *S*-BTBT-CONHR thin films on HOPG with low surface roughnesses of 0.10, 0.14, 0.11, and 0.18 nm measured in (a), (b), (c), and (d), respectively.

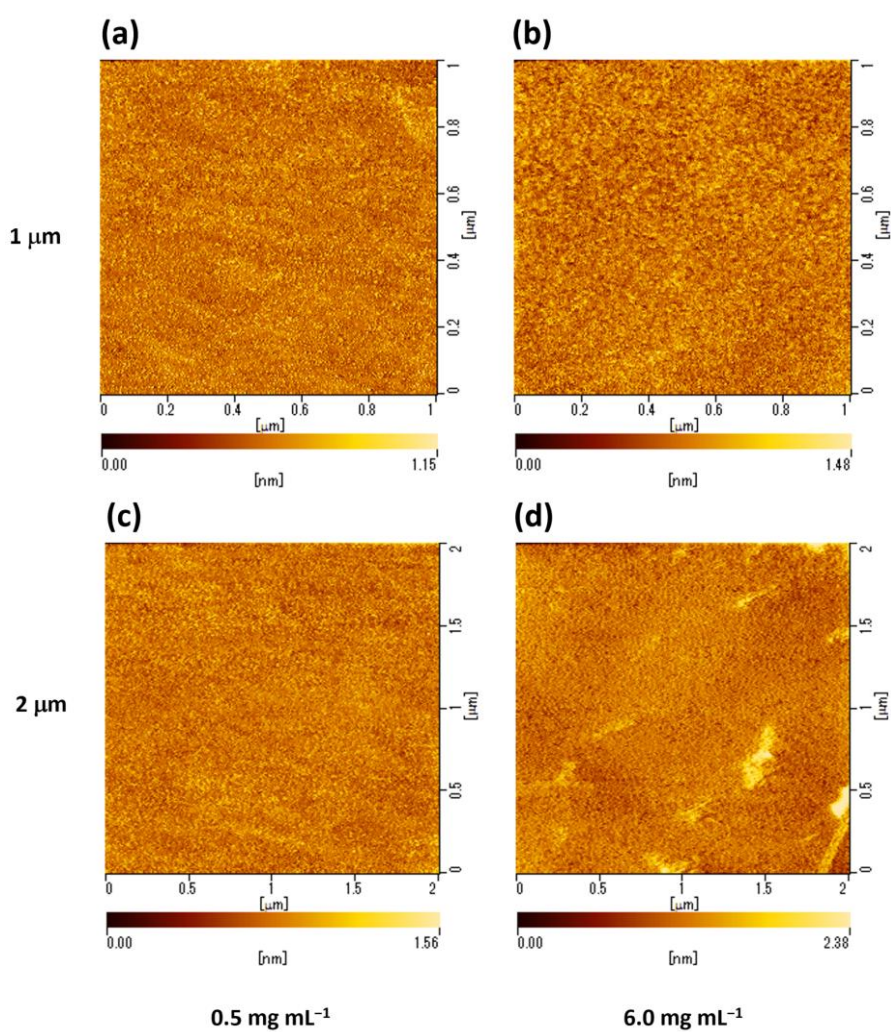

**Fig. S9** AFM images of the *S*-BTBT-CONHR thin films on HOPG spin-coated from the solutions with concentrations of (a, c) 0.5 and (b, d) 6.0 mg mL$^{-1}$, respectively.

**Estimation of the chiral film thickness on HOPG substrate.** We attempted to measure the thickness of the chiral films. In this measurement, we used an AFM cantilever for scratching the chiral film during AFM scanning with a contact mode in a scanning area of 1 × 1 μm², and measured depth. As a result, the thickness is obtained as 3–4 nm in the chiral film spin-coated with a solution of 6 mg mL$^{-1}$. In the chiral film spin-coated with a solution of 0.5 mg mL$^{-1}$, we could not obtain the clear depth profile, indicating that the thickness may be as thin as ~1 nm. Here, to estimate the thickness, we have performed the fitting calculation using Simmons' equation.[2] According to Simmons' equation, the tunneling current *I* through the barrier is expressed as

$$I = \frac{eA}{2\pi hd^2}\left[\left(\phi - \frac{eV}{2}\right)\exp\left(-\frac{4\pi d}{h}\sqrt{2m\left(\phi - \frac{eV}{2}\right)}\right) - \left(\phi + \frac{eV}{2}\right)\exp\left(-\frac{4\pi d}{h}\sqrt{2m\left(\phi + \frac{eV}{2}\right)}\right)\right] \quad (S1)$$

where *e* is the electron charge, *m* is the electron mass, *h* is Planck's constant, *d* is the barrier thickness, $\phi$ is the barrier height, *A* is the contact area, and *V* is the bias voltage. In this fitting calculation, we considered that $\phi$ between the chiral molecule and HOPG is the same as that between the chiral molecule and CoPtCr because the measured *I–V* curves are symmetric. We used *d*, $\phi$, and *A* as fitting parameters. Here, we assumed that there is no change in *d* and *A* between up and down magnetization. As shown in Fig. S10, the fitting results of up (down) magnetization are in good agreement with the experimental data when the fitting parameters are *d* = 1.19 (1.19) nm, $\phi$ = 2.75 (2.04) eV, and *A* = 3.11 (3.11) × 10$^{-15}$ m². The fitting results indicate that the thickness is estimated to be approximately 1 nm.

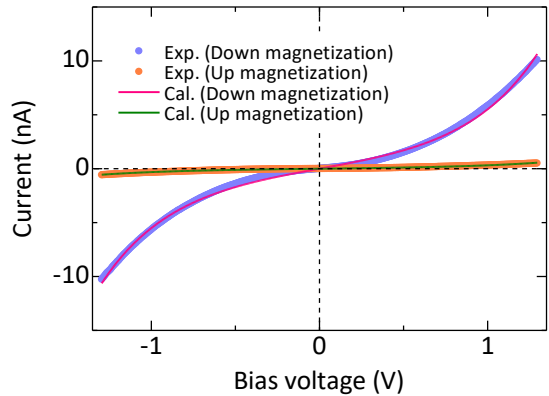

**Fig. S10** Fitting results of *I–V* curves by Simmons' equation in the *S*-BTBT-CONHR thin films on HOPG substrates spin-coated from a solution of 0.5 mg mL$^{-1}$.

**Raw data of *I–V* curves measured by mc-AFM.** Fig. S11 shows the raw data of forty *I–V* curves measured with a cantilever which is magnetized along the up or down direction in the *S*-BTBT-CONHR thin films on HOPG substrates spin-coated from the solutions with concentrations of 0.5 and 6.0 mg mL$^{-1}$. In addition, we attempted to add the error bar on the averaged *I–V* plots. Since each *I–V* curve includes more than 500 plots, the error bar is not clear. Here, we show the *I–V* plots with the error bar which consists of 1/10 plots from the original averaged data (Fig. S12) to make the error bar visible.

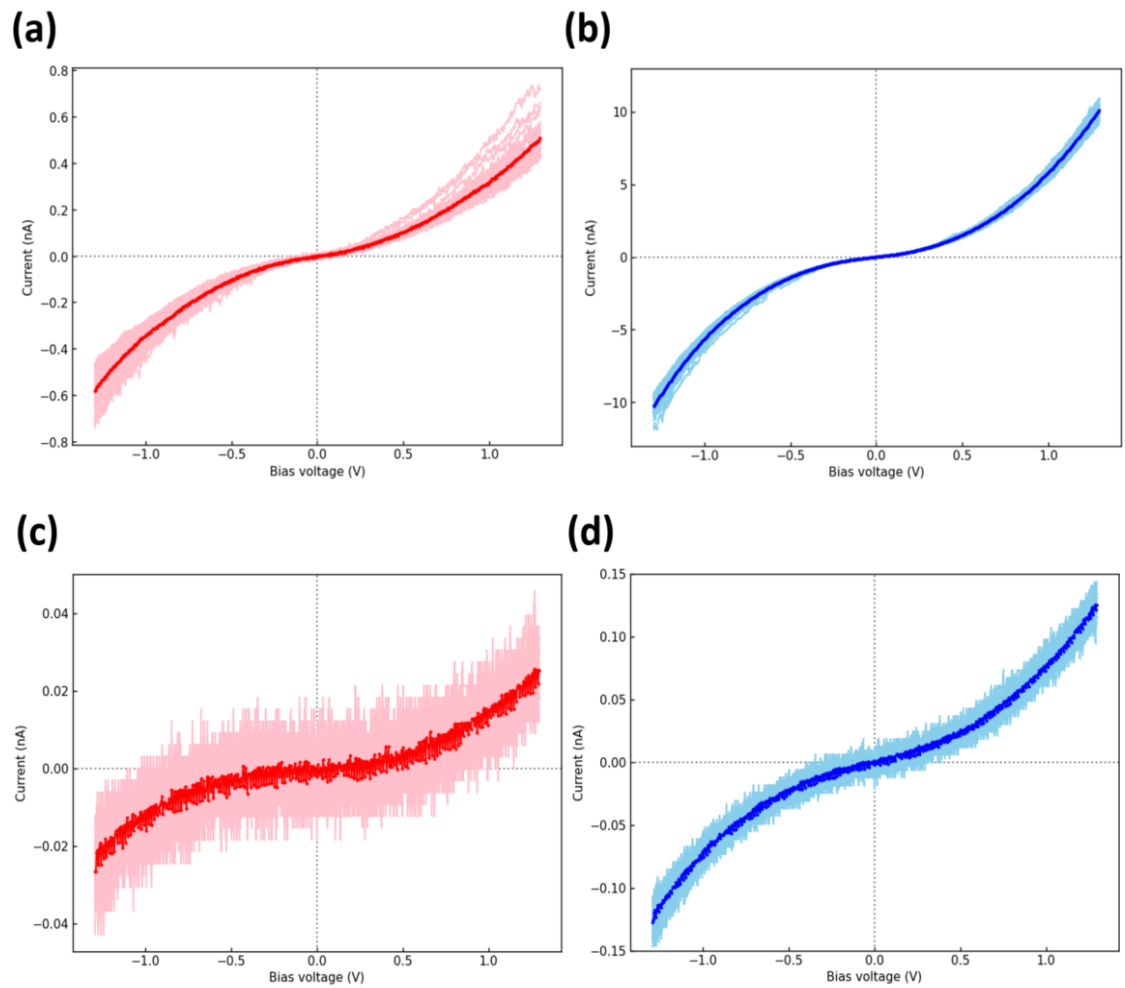

**Fig. S11** Raw data of forty *I–V* curves measured with a cantilever which is magnetized along the up or down direction in the *S*-BTBT-CONHR thin films on HOPG substrates spin-coated from the solutions with concentrations of (a, b) 0.5 and (c, d) 6.0 mg mL$^{-1}$.

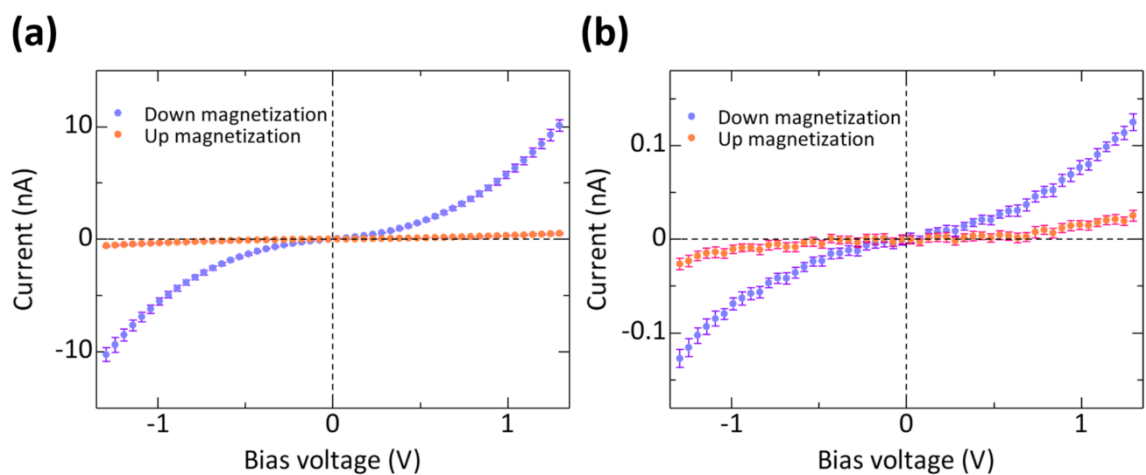

**Fig. S12** The averaged *I–V* plots with the error bar which consists of 1/10 plots from the original averaged data in the *S*-BTBT-CONHR thin film spin-coated using solutions with concentrations of (**a**) 0.5 and (**b**) 6.0 mg mL$^{-1}$, respectively.

**Magnetization curves of the Ni$_{78}$Fe$_{22}$ thin films.** Fig. S13 shows the magnetization curves measured in the Ni$_{78}$Fe$_{22}$ thin films before thermal pressing. The coercivity of the Ni$_{78}$Fe$_{22}$ thin films is 4 Oe.

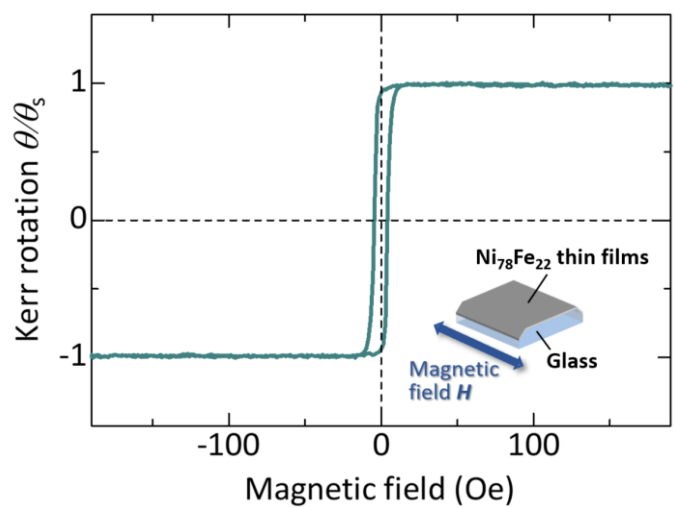

**Fig. S13** Magnetization curves of the Ni$_{78}$Fe$_{22}$ thin films before thermal pressing.

**Additional MR curves.** In addition to Au/*S*-BTBT-CONHR/Ni$_{78}$Fe$_{22}$ nanodevices, we have fabricated Au/Ni$_{78}$Fe$_{22}$ nanodevices without a chiral molecular layer. These nanodevices do not show the MR effect with magnetic hysteresis as shown in Fig. S14, which is the evidence that the MR effect observed in Fig. 6(b) is attributed to the CISS effect. The two weak peaks at a magnetic field of approximately ±40 Oe originate from anisotropic MR (AMR) of the Ni$_{78}$Fe$_{22}$ electrode. This result indicates that when the device is short-circuited, the resistance is lower than that in the device with sandwiching molecules. This is presented in our previous study.[1] Normally, the similar resistance values are obtained in the devices which are fabricated by the same conditions.[1] Therefore, we can distinguish whether the device is short-circuited or not by the resistance value. In addition, in our fabrication method, we can fabricate chiral molecular layers without any interdiffusion of Au atoms because the top Au layer is not deposited on molecules by thermal evaporation or sputtering method. Therefore, Au atoms cannot penetrate easily into molecules.

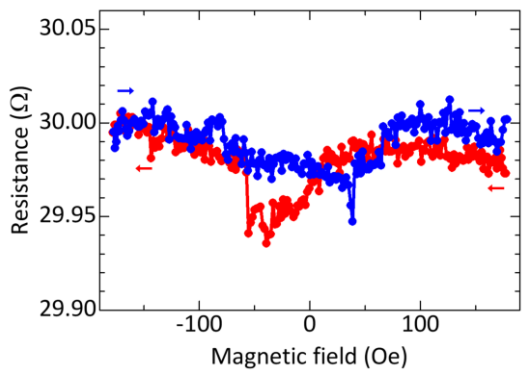

**Fig. S14** MR properties in Au/Ni$_{78}$Fe$_{22}$ nanodevices.

In the main text, a CISS-based MR effect is observed at a constant current of 40 µA in Fig. 6(b). Fig. S15 shows the MR effect in the Au/*S*-BTBT-CONHR/Ni$_{78}$Fe$_{22}$ nanodevices at different currents of 400, −60, and −120 µA. This means that we can observe CISS-based MR at various current. Here, the MR curves show a negative peak under a magnetic field from −60 to −40 Oe, originating from the AMR effect of the Ni$_{78}$Fe$_{22}$ electrode with a coercivity of 39 Oe. It has not been clarified that the reason why there is no or a small negative peak at positive magnetic field. Estimated from the AMR effect, the magnetization rotates under positive magnetic fields more easily than under negative fields. The asymmetric phenomenon of the magnetic anisotropy may be related to the chiral molecular layer. Specifically, chiral molecules might affect the magnetic anisotropy in its adjacent magnetic layer. Here, the device resistance at a positive current is larger than that at a negative current. This behavior can originate from the work function difference between the Au and $Ni_{78}Fe_{22}$ electrodes. In *S*-BTBT-CONHR, the HOMO and LUMO are calculated as −5.706 and −1.617 eV, respectively, whereas the work functions of Au or $Ni_{78}Fe_{22}$ electrodes are 5.4 and 5.0 eV.[3,4] These pictures imply that the hole injection barrier is higher at the $Ni_{78}Fe_{22}$/*S*-BTBT-CONHR interface than that at the Au/*S*-BTBT-CONHR interface.

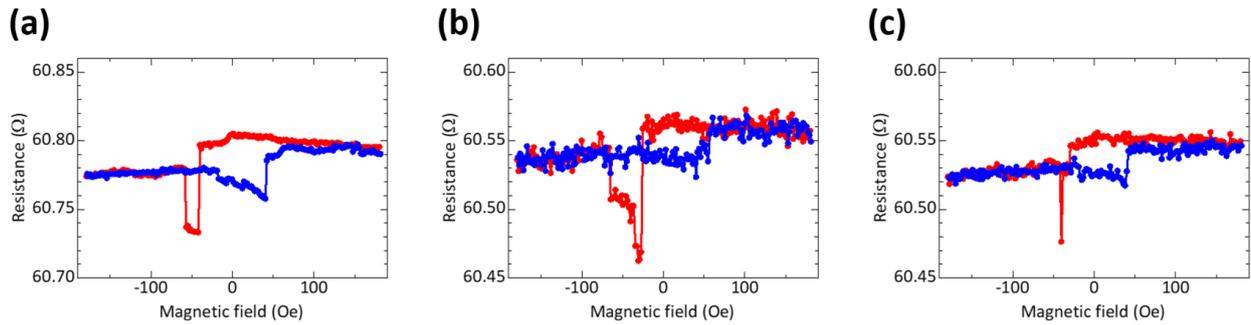

**Fig. S15** MR curves in Au/*S*-BTBT-CONHR/$Ni_{78}Fe_{22}$ nanodevices at room temperature. The currents are (a) 400, (b) −60, and (c) −120 μA.

We also fabricated the Au/*racemic*-BTBT-CONHR/$Ni_{78}Fe_{22}$ device, in which the one-side electrode is coated with *racemic*-BTBT-CONHR. In this device, as shown in Fig. S16, the inverse MR effect is observed with respect to the MR polarity of the *S*-BTBT-CONHR device. Therefore, these results indicate that the observed MR originates from the CISS effect of chiral BTBT-CONHR. In addition, as we attempt to synthesize *R*-derivative, it requires two or three months and the use of expensive reagents and special experimental equipment. From the viewpoint of research efficiency and effective use of resources, we consider that it is preferable to develop our research based on the data already obtained for the *S*-derivative (and *racemic* mixture) rather than to synthesize and evaluate the *R*-derivative at this stage. In particular, it is considered important to focus on elucidating the mechanisms of interesting phenomena observed in the *S*-derivative and exploring the potential for applications. If in the future it is suggested that there may be unexpected differences between counterparts in physical properties, we synthesize and evaluate the *R*-derivative to be considered at that time. Thus, in this study, we conclude that our MR results can originate from the CISS effect with the experimental results using *S*-BTBT-CONHR (and the observed MR effect in the Au/*racemic*-BTBT-CONHR/Ni$_{78}$Fe$_{22}$ device). Here, we attempted to fabricate the nanodevice with a thinner chiral film than that in our original experiments. To realize such a device, the chiral film was spin-coated on only one electrode, which means that the chiral film thickness is half of that in the original device. As shown in Fig. S17, we have successfully observed the MR effect in the device with the thinner chiral film.

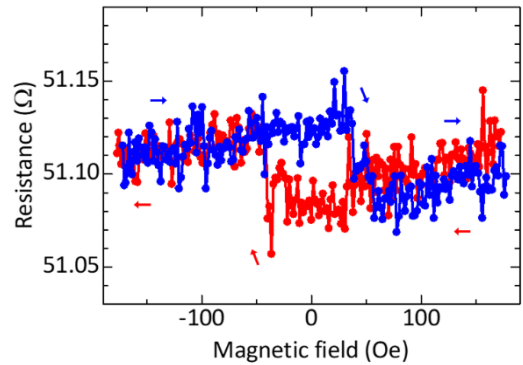

**Fig. S16** MR effect observed in the Au/*racemic*-BTBT-CONHR/Ni$_{78}$Fe$_{22}$ device.

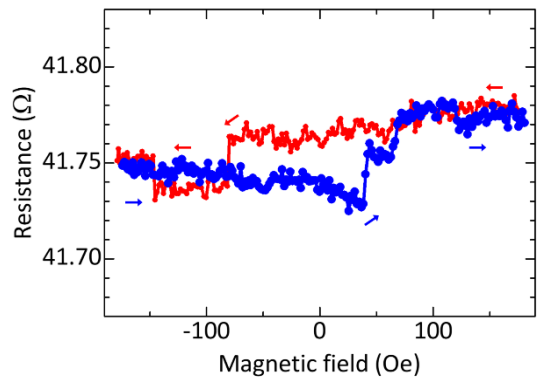

**Fig. S17** MR effect observed in the Au/*S*-BTBT-CONHR/Ni$_{78}$Fe$_{22}$ device in which a chiral film is spin-coated onto only one electrode.